\documentclass[aps,prc, twocolumn]{revtex4-2}
\usepackage{graphicx}
\usepackage{amsmath}
\usepackage{amsfonts}
\usepackage{comment}
\begin{document}

\title{Influence of the residual magnetic field on the azimuthal distribution of final-state particles in photon-nuclear processes}
\author{Zhan Zhang}
\affiliation{State Key Laboratory of Particle Detection and Electronics, University of Science and Technology of China, Hefei 230026, China}
\author{Xin Wu}
\email[Corresponding author, ]{Xin Wu Address: No. 96 Jinzhai Road, Hefei city, Email: wuxinust@mail.ustc.edu.cn}
\affiliation{State Key Laboratory of Particle Detection and Electronics, University of Science and Technology of China, Hefei 230026, China}
\author{Xinbai Li}
\affiliation{State Key Laboratory of Particle Detection and Electronics, University of Science and Technology of China, Hefei 230026, China}
\author{Wangmei Zha}
\email[Corresponding author, ]{Wangmei Zha Address: No. 96 Jinzhai Road, Hefei city, China Tel: +86 551 63607940  Email: first@ustc.edu.cn}
\affiliation{State Key Laboratory of Particle Detection and Electronics, University of Science and Technology of China, Hefei 230026, China}
\author{Zebo Tang}
\affiliation{State Key Laboratory of Particle Detection and Electronics, University of Science and Technology of China, Hefei 230026, China}

\begin{abstract}
In relativistic heavy-ion collisions, charged particles are accelerated to nearly the speed of light, and their external electromagnetic fields can be effectively approximated as quasi-real photons. These photons interact with another nucleus via photon-nuclear interactions, producing vector mesons. These vector mesons possess extremely low transverse momentum ($p_T\sim0.1$ GeV/$c$), distinguishing them from particles produced via hadronic interactions. STAR and ALICE have observed $J/\psi$, $\rho^0$ and other vector mesons with very low $p_T$, which are well described by photoproduction models. This unique characteristic of having extremely low transverse momentum allows them to serve as a novel experimental probe. Recent STAR results show that the equivalent photons in photoproduction processes are fully linearly polarized, affecting the azimuthal distribution of final-state particles like
$\rho^0 \rightarrow \pi^+ \pi^-$. Since the polarization links to the initial collision geometry, the $\rho^0$ azimuthal modulation can probe nuclear structure. However, the post-collision magnetic field may deflect these particles, distorting the azimuthal distribution and complicating structure measurements. We simulated the distribution of residual magnetic fields over time under different collision conditions using UrQMD for Au+Au collisions at $\sqrt{s_{NN}}=200$ GeV and calculated their effects on the azimuthal modulation ($\left<cos2\phi\right>$) of photoproduced $\rho^0$. Our results show that in peripheral collisions, the field significantly alters the $\left<cos2\phi\right>$ for photoproduced $\rho^0$ with $p_T\approx$ 0.1 GeV$/c$. This provides key insights for future nuclear structure studies via photoproduction in peripheral collisions.

\end{abstract}
\maketitle
\section{Introduction}

In recent years, relativistic heavy-ion collision experiments have been recognized as a new platform for studying nuclear structure, such as utilizing the anisotropic flow generated in central collisions to determine nuclear structure \cite{ma2023influence, abdallah2022evidence, xu2018importance}. Meanwhile, photoproduction processes have been attracting increasing attention and are considered a new probe for investigating nuclear structure \cite{lin2025nuclear, star2023tomography}. In relativistic heavy-ion collisions, as charged particles are accelerated to nearly the speed of light, the electromagnetic fields generated by the nuclei are highly Lorentz-contracted, resulting in complete linear polarization \cite{ krauss1997photon, bertulani1988electromagnetic}. The scattering of linearly polarized photons with nuclei produces vector mesons (e.g., $\rho^0, \omega, J/\Psi$), which will inherit the polarization state of the initial photons \cite{criegee1968rho,PhysRevLett.24.1364,Eisenberg:1975jr,ZEUS:1996esk,STAR:2007elq,LEPS:2017nqz}. The decays of these vector mesons will tend to align with the direction of the polarization vector, inducing a specific cosine modulation in the angular distribution of the final-state particles~\cite{zha2021exploring,Xing:2020hwh}. In recent measurements of photoproduced $\rho^0$ mesons, STAR has clearly observed this cosine modulation phenomenon \cite{star2023tomography}. On the other hand, since the polarization vector is highly sensitive to the initial geometry (impact parameter, nuclear size, charge distribution, etc.), the experimentally measured cosine modulation can be used to extract the shape of the atomic nucleus \cite{zha2021exploring, xing2020cos, alice2024measurement}. By utilizing polarized photons, this nuclear imaging technique addresses the charge-distribution-only constraint inherent in conventional electron scattering experiments, thereby providing a novel and robust means for nuclear structure research. Alternatively, by measuring the correlation between this azimuthal modulation and the event plane, one can determine the impact parameter \cite{wu2022reaction}.

However, it is noteworthy that since electromagnetic fields act as retarded potentials, residual magnetic fields persist in the particle production region even after the collision \cite{kharzeev2008effects, deng2012event}. These fields deflect the trajectories of final-state particles, thereby distorting the experimentally measured azimuthal modulation patterns~\cite{Toneev:2012zx,Chen:2024aom}.
Consequently, this effect compromises the accuracy of extracting geometrical information of the collision system using such observables. In this paper, we calculate the coordinate-space and momentum-space distribution of photoproduced $\rho^0$ mesons in Au+Au collisions at 200 GeV and simulate their isotropic decays into $\pi^+ \pi^-$ pairs. Using the UrQMD model \cite{bass1998microscopic, bleicher1999relativistic}, we simulate Au+Au collisions across different centrality classes and compute the residual magnetic fields generated by strongly interacting particles. We then apply these residual magnetic fields to the $\pi^+ \pi^-$ pairs and quantify the resulting cosine modulation induced by these field effects. Our calculations reveal that in peripheral collisions, the abundant particle production in the interaction region leads to a significantly slower decay of the residual magnetic field and markedly alters its spatial distribution compared to ultra-peripheral collisions. Consequently, this effect produces a pronounced impact on the azimuthal modulation pattern in such collision systems. Our study provides a new theoretical reference for future experimental investigations that aim to determine the initial geometry using polarized photons in peripheral collisions.  Although this paper primarily investigates the influence of the residual magnetic field on the angular distribution of final-state particles in photon-nucleus interactions, such effects should also impact the angular distribution of particles produced via photon-photon interactions\cite{Li:2019sin,Klein:2018fmp}. Therefore, corresponding contributions must be considered in related studies.

\section{Methodology}
The computational approach adopted in this paper is as follows: First, we calculate the coordinate-space and momentum-space distribution of photoproduced $\rho^0$ mesons. The $\rho^0$ mesons are sampled from these distributions and decay isotropically (without considering polarization effects, focusing solely on the potential influence of the residual magnetic field), with their finite lifetime properly considered. Next, we use the UrQMD model and Electrodynamics to compute the spatial and temporal evolution of the magnetic field for different collision centralities. Finally, we apply the magnetic field to the $\pi^+\pi^-$ pairs from $\rho^0$ decay and calculate the second-order cosine modulation coefficient of their final-state azimuthal distribution. In this study, the right-handed coordinate system for the vector meson decay is constructed such that the $z$-axis points along the vector meson's flight direction in the photon-nucleon center-of-mass frame, the $y$-axis is normal to the photoproduction plane, and the $x$-axis is given by $y \times z$ to complete the right-handed system, $\phi$ is the azimuthal angle. 

\subsection{The coordinate distribution of photoproduced $\rho^0$}

The calculations in this section primarily employ the Equivalent Photon Approximation (EPA)+Vector Meson Dominance (VMD) model. The electromagnetic field generated by relativistically moving nuclei can be effectively described as equivalent photons, with the photon flux given by the EPA model\cite{von1934radiation, williams1934nature} as:
\begin{equation}
    \begin{aligned}
        \frac{d^3N_\gamma({\omega}_{\gamma},\vec{x}_{\perp})}{d{\omega}_{\gamma}\,d\vec{x}_{\perp}}=\frac{4Z^2\alpha}{{\omega}_{\gamma}}{\left |\int\frac{d^2\vec{k}_{\gamma\perp}}{(2\pi)^2}\vec{k}_{\gamma\perp}\frac{F_\gamma(\vec{k}_\gamma)}{\lvert\vec{k}_\gamma\rvert^2}e^{i\vec{x}_\perp\cdot{\vec{k}_{\gamma\perp}}}\right |}^2,
        \\
        \vec{k}_\gamma=\bigg(\vec{k}_{\gamma\perp},\frac{\omega_\gamma}{\gamma_c}\bigg),\ \omega_\gamma=\frac{1}{2}M_V e^{{\pm}y},
    \end{aligned}
    \label{photonflux}
\end{equation}
where \(\vec{x}_\perp\) is the photon's position, \(\vec{k}_{\gamma\perp}\) is its transverse momentum, \(\omega_\gamma\) is the photon's energy, \(\gamma_c\) is the Lorentz factor of the photon-emitting nucleus, \(Z\) is the electric charge of the nucleus, \(\alpha\) is the electromagnetic coupling constant, \(M_V\) and \(y\) are the mass and rapidity of the vector meson. \(F_\gamma(\vec{k}_\gamma)\) represents the nuclear electromagnetic form factor, obtained through Fourier transformation of the nuclear charge distribution. Here we use the Woods-Saxon distribution to describe the shape of Au nuclear:
\begin{equation}
    \rho_A(r)=\frac{\rho_0}{1+\exp{[(r-R_{WS})/d]}},
\end{equation}
where $R_{WS}=6.38$ fm is the nuclear radius, $d=0.535$ fm is the skin depth \cite{barrett1977nuclear}, and \(\rho_0\) is the normalization factor.

The scattering amplitude distribution in coordinate space \(\Gamma_{\gamma A\rightarrow\rho^0A}(\vec{x}_\perp)\) with nuclear shadowing effects can be computed using the Glauber model \cite{miller2007glauber} and VMD approach \cite{bauer1978hadronic}:

\begin{equation}
    \Gamma_{\gamma A\rightarrow\rho^0A}
(\vec{x}_\perp)=\frac{f_{\gamma N\rightarrow\rho^0N}(0)}{\sigma_{\rho^0N}}2\bigg[1-\exp\bigg(-\frac{\sigma_{\rho^0N}}{2}T^\prime(\vec{x}_\perp)\bigg)\bigg],
\end{equation}
where $f_{\gamma p\rightarrow \rho^0 N}(0)$ is the parametrized forward-scattering amplitude, $\sigma_{\rho^0 N}$ is the total cross section for $\rho^0 N$ scattering, given by
\begin{equation}
    \sigma_{\rho^0N}=\frac{f_V}{4\sqrt{\alpha}C}f_{\gamma N\rightarrow\rho^0N},
\end{equation}
where $f_V$ is the coupling factor of vector meson and photon~\cite{Klein:1999qj}, C is a correction factor for the off-diagonal diffractive interaction~\cite{Hufner:1997jg}. $T^\prime(\vec{x}_\perp)$ is the modified thickness function accounting for the coherence length effect:
\begin{equation}
    T^\prime(\vec{x}_\perp)=\int_{-\infty}^{+\infty}\rho_A(\sqrt{|\vec{x}_\perp|^{\,2}+z^2})e^{iq_Lz}dz,\ q_L=\frac{M_V e^y}{2\gamma_c},
\end{equation}
The production amplitude is expressed by convoluting the photon flux amplitude with the scattering amplitude $\Gamma_{\gamma A\rightarrow\rho^0A}(\vec{x}_\perp)$ \cite{zha2018coherent}:
\begin{equation} \label{eq:9}
    \vec{A}(\vec{x}_\perp)=(A_x,A_y)=\sqrt{\frac{d^3N_\gamma({\omega}_{\gamma},\vec{x}_{\perp})}{d{\omega}_{\gamma}\,d\vec{x}_{\perp}}}\Gamma_{\gamma A\rightarrow\rho^0A}(\vec{x}_\perp).
\end{equation}
The momentum-space production amplitude is derived through a Fourier transform of its coordinate-space counterpart:
\begin{equation} \label{eq:12}
    \vec{A}(\vec{p}_\perp)=\frac{1}{2\pi}\int d^2\vec{x}_\perp(\vec{A}_1(\vec{x}_\perp)+\vec{A}_2(\vec{x}_\perp))e^{i\vec{p}_\perp\cdot\vec{x}_\perp},
\end{equation}
where \(\vec{A}_1(\vec{x}_\perp)\) and \(\vec{A}_2(\vec{x}_\perp)\) are the spatial amplitude distributions in the transverse plane for the two nuclei. 

In experiments, the photoproduction process is typically accompanied by nuclear excitation with neutron emission induced by photons. We calculate the probability of this process by multiplying the photon flux with the photon absorption cross section. The lowest-order probability for nuclear excitation accompanied by single or multiple neutron emission ($X_n$) can be expressed as \cite{zha2021exploring}:
\begin{equation}
    m_{X_n}(b)=\int d\omega n(\omega,b)\sigma_{\gamma A\rightarrow A^*}(\omega),
\end{equation}
where $n(\omega,b)$ is the photon flux described by q.\ref{photonflux} and $\sigma_{\gamma A\rightarrow A^*}(\omega)$ is the photon-excitation cross section obtained from experimental data \cite{veyssiere1970photoneutron, lepretre1981measurements, carlos1984total, armstrong1972total, caldwell1973total, michalowski1977experimental, armstrong1972total2}. The $X_n X_n$ probability (mutual dissociation with $\geq1$ neutron for each nucleus) is then:
\begin{equation}
    P_{X_n X_n}(b)=(1-\text{exp}[-m_{X_n}(b)])^2.
\end{equation}

\begin{figure}
    \centering
    \includegraphics[width=0.8\linewidth]{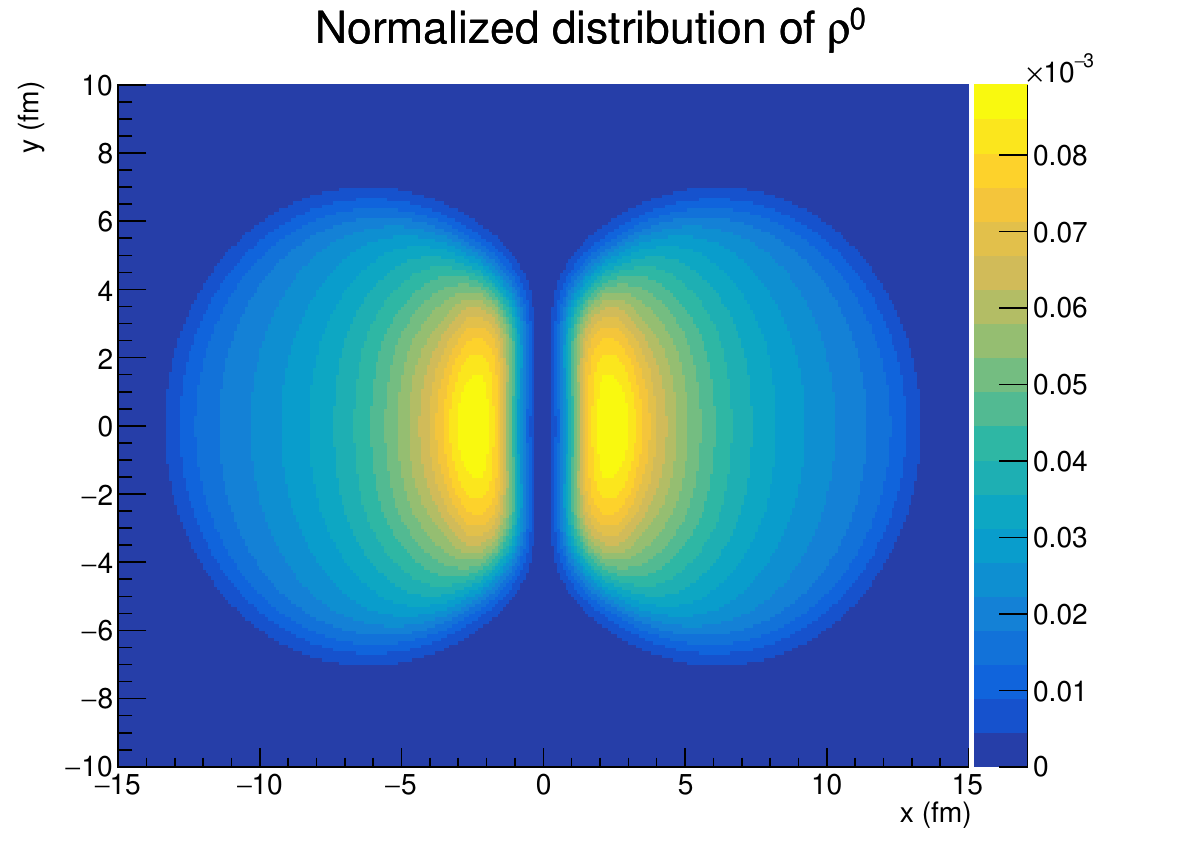}
    \caption{The photoproduced $\rho^0$ distribution in coordinate space for Au+Au collisions at $\sqrt{s_{NN}}=200$ GeV with $b=10$ fm.}
    \label{rho0xydis}
\end{figure}

\begin{figure}
    \centering
    \includegraphics[width=0.8\linewidth]{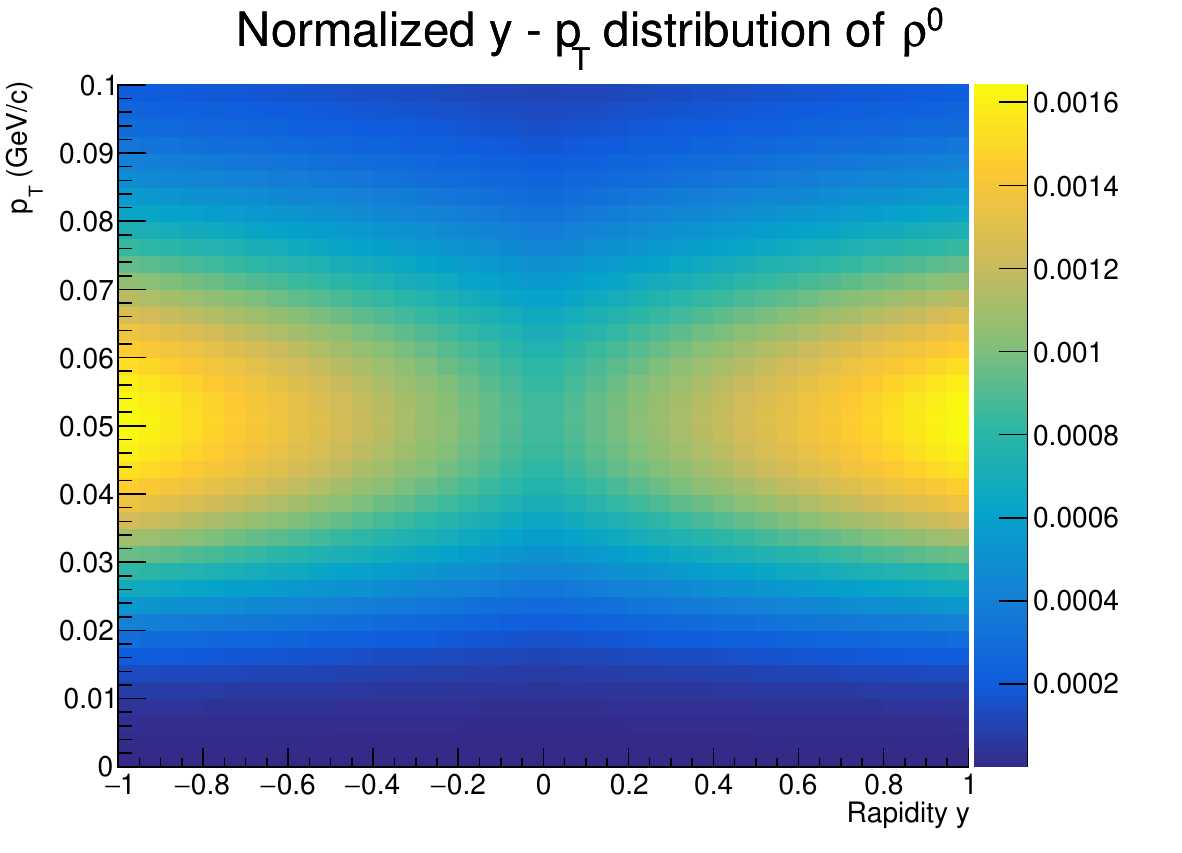}
    \caption{The rapidity-$p_T$ distribution of photorpoduced $\rho^0$ for Au+Au collisions at $\sqrt{s_{NN}}=200$ GeV with $b=10$ fm. The peak appears at a position around $p_T\approx0.05$ GeV/$c$. }
    \label{rho0ptydis}
\end{figure}

Figure~\ref{rho0xydis} presents the coordinate-space distribution of photoproduced $\rho^0$ mesons in $\sqrt{s_{NN}}=200$ GeV Au+Au collisions at an impact parameter of 10 fm, calculated following the aforementioned computational procedure. The two nuclei can be treated as finite-sized slits that emit $\rho^0$ wave functions, which then undergo coherent superposition. Consequently, distinct interference patterns and diffraction rings emerge in momentum space, producing oscillatory peak-and-valley structures in the transverse momentum spectrum, as clearly demonstrated in Fig.~\ref{rho0ptydis}. We sample the position, momentum and rapidity information of $\rho^0$ mesons from these distributions and let them decay isotropically into $\pi^+\pi^-$ pairs with an exponential time distribution ($\frac{1}{\tau} e^{-\tau/t}$, $\tau$ is the lifetime of $\rho^0$ meson) for subsequent calculations.


\subsection{The calculation of residual magnetic fields}
The calculation of the residual magnetic field is divided into two scenarios: the UPC case without hadronic interaction and the peripheral collision (PC) case with hadronic interaction. 

The magnetic field strength induced by a nucleus at a position $\vec{x}$ and time t is defined by the Lienard-Wiechert potentials~\cite{Skokov:2009qp}:
\begin{equation}
e\vec{B}\left(t,\vec{x}\right)=\alpha_{EM}\sum\limits_{n=1}^{N} Z_n\frac{1-v_n^2}{\left(R_n-\vec{R_n}\vec{v_n}\right)^3}\left[\vec{v_n}\times\vec{R_n}\right],
\label{LWpotential}
\end{equation}
where $\alpha_{EM}$ is the fine-structure constant, $Z_n$ is the electric charge of the n-th particle and $\vec{R_n}=\vec{x}-\vec{x_n}$ is the radius vector of the n-th particle, $\vec{x_n}$ and $\vec{v_n}$ represent the position and velocity of the n-th particle, respectively. For UPC case, where the impact parameter is typically large and the nuclei remain intact without fragmentation, the Au nucleus can be approximated as point-like particles. The magnetic field at $z=0$ is:
\begin{equation}
e\vec{B}\left(t,\vec{x}\right)=\frac{\alpha_{EM}Z_{Au}\gamma \beta}{\left(\gamma^2\beta^2t^2+|\vec{x}|^2\right)^{3/2}}.
\end{equation}
The total magnetic field is the sum of the magnetic field induced by the two colliding Au nuclei.

\begin{figure}
    \centering
    \includegraphics[width=1.0\linewidth]{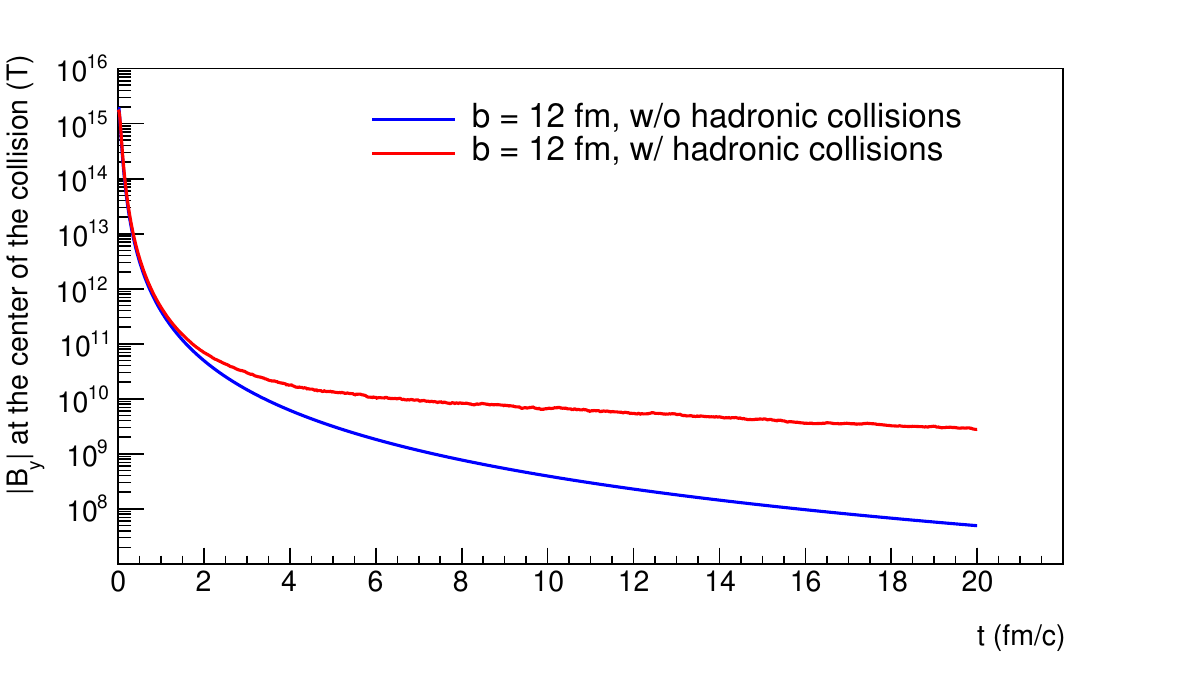}
    \caption{Magnetic field variation over time after the collision, where $t=0$ marks the collision time. The blue curve represents the case without hadronic interaction, while the red curve corresponds to the case with hadronic interaction.}
    \label{B_time}
\end{figure}

For PC case, since the atomic nuclei have collided and are no longer intact, we use UrQMD to simulate the post-collision particle distributions and Eq.~\ref{LWpotential} to calculate the magnetic field distribution. Figure~\ref{B_time} shows the variation of the magnetic field over time after the collision, where $t=0$ fm/$c$ marks the collision time. The blue curve represents the UPC case, while the red curve represents the PC case. It can be seen that the magnetic field decay in the PC case is significantly slower than in the UPC case. This is because the particles produced in the collision remain in the interaction region, providing additional contributions to the magnetic field. 

\begin{figure}
    \centering
    \includegraphics[width=1.0\linewidth]{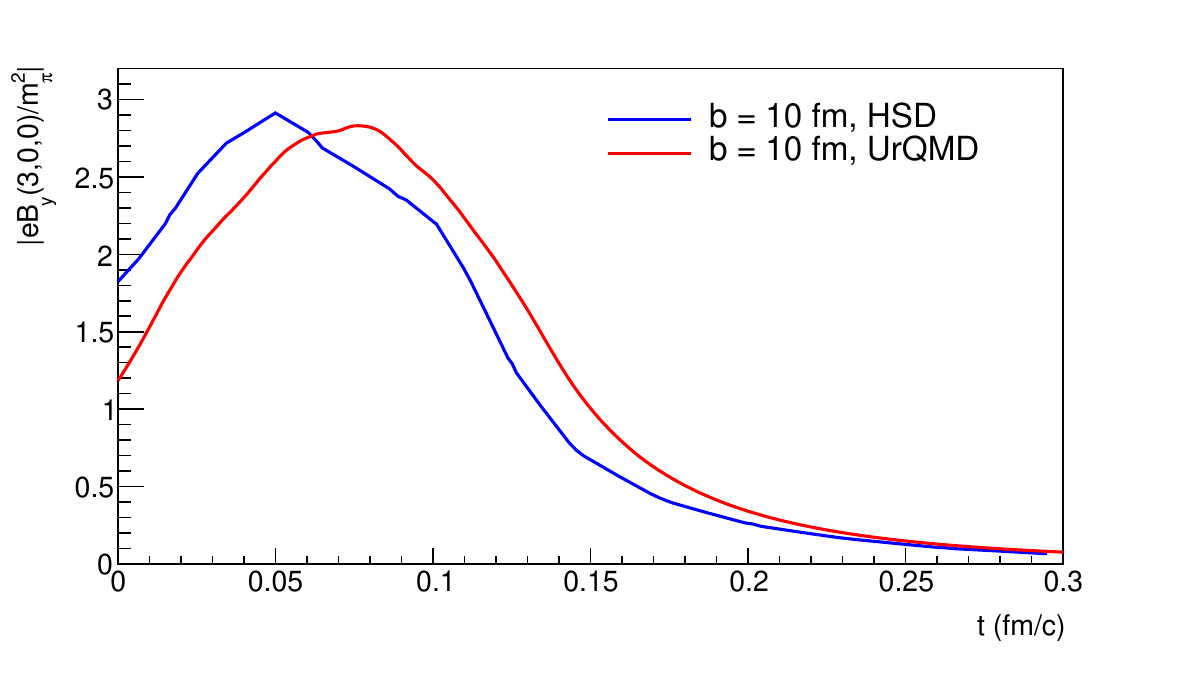}
    \caption{The $|B_y|$ at $x=3$ fm and $y=0$ fm calculated from UrQMD model (red curve) compared with that calculated from HSD model (blue curve)~\cite{Voronyuk:2011jd}. The different models yield comparable magnetic field strengths and similar temporal evolution patterns.}
    \label{B_compare}
\end{figure}

In addition to UrQMD, there are other models that can describe the motion of particles produced after relativistic heavy-ion collisions~\cite{Bleicher:2022kcu}. We compared the magnetic fileds calculated by the UrQMD model with those from the HSD model~\cite{Voronyuk:2011jd}, shown as Fig. ~\ref{B_compare}. It can be observed that at $b=10$ fm,  $x=3$ fm, $y=0$ fm (a position located within the primary photoproduction region of $\rho^0$, see Fig.~\ref{rho0xydis}), the $|B_y|$ generated from both models exhibit comparable magnitudes and temporal evolution patterns, consequently leading to similar deflection effects on the $\pi^+\pi^-$ pairs originating through the photoproduction and subsequent decay of $\rho^0$. In the present work, we will focus primarily on the results obtained from the UrQMD model calculations.

While the contribution from the electric field should in principle be accounted for, its opposing effects on positively and negatively charged particles result in negligible influence on 2$\left<cos2\phi\right>$. This was explicitly verified by simulating $\pi^+\pi^-$ evolution in the residual electric field, which consistently yielded null results for 2$\left<cos2\phi\right>$. We therefore disregard them hereafter.
\section{Results}
\begin{figure}
    \centering
    \includegraphics[width=1.0\linewidth]{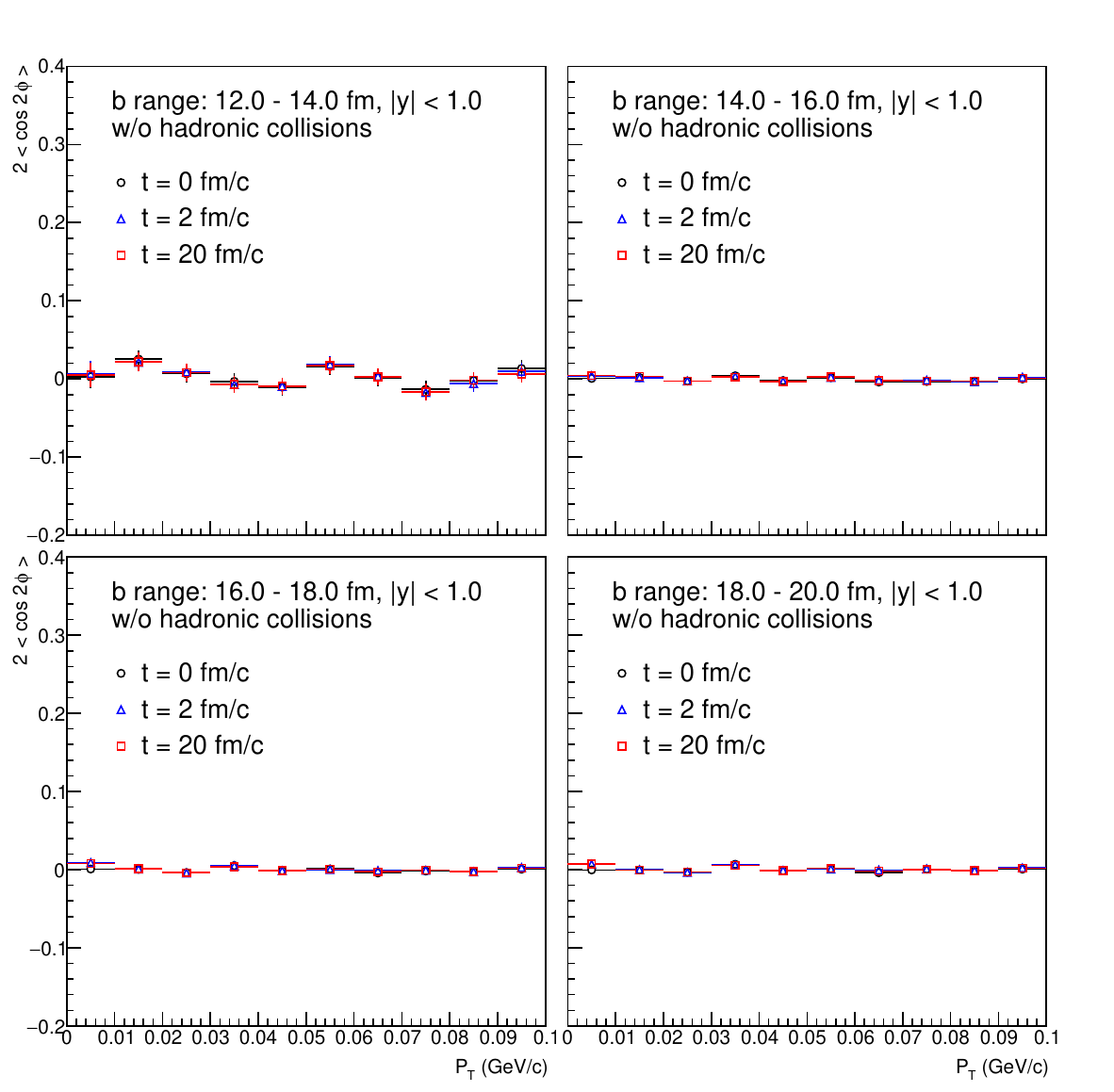}
    \caption{The 2$\left<cos2\phi\right>$ induced by the residual magnetic field in UPC case as a function of the $\pi^+\pi^-$ pair $p_T$. Different panels represent different b ranges. The black circles represent the results at $t=0$ fm/$c$, the blue triangles at $t=2$ fm/$c$, and the red squares at $t=20$ fm/$c$.}
    \label{cos2phi_pt_UPC}
\end{figure}

We investigate the azimuthal modulation (2$\left<cos2\phi\right>$) of final-state particles by applying the magnetic field to the $\pi^+ \pi^-$ decays from the photoproduced $\rho^0$ mesons which are sampled according to the calculated distribution described in Section II. A, where $\phi$ is defined in $\rho^0$ rest frame as $\phi^{\pi^++\pi^-}-\phi^{\pi^+-\pi^-}$. We selected three characteristic time points to monitor the evolution of 2$\left<cos2\phi\right>$: (1) $t=0$ fm/$c$, corresponding to the instant of nuclear collision with initial $\rho^0$ photoproduction. (2) $t=2$ fm/$c$,when most $\rho^0$ mesons have decayed and no additional particles are produced in the nuclear overlap region (particularly for peripheral collisions). (3) $t=20$ fm/$c$, representing the final-state configuration where the decayed $\pi^+\pi^-$ pairs have propagated sufficiently through the magnetic field. Figure~\ref{cos2phi_pt_UPC} shows the 2$\left<cos2\phi\right>$ induced by the residual magnetic field in UPC case as a function of $p_T$ for different impact parameter ranges. It can be seen that for the UPC case, 2$\left<cos2\phi\right>$ does not show significant changes in different b ranges with $p_T$ and time. This is because, as shown by the blue curve in Fig.~\ref{B_time}, the magnetic field generated after the "collisions" of the two Au nuclei decays rapidly and is primarily concentrated near the origin, where very few $\rho^0$ are produced. Consequently, it does not exhibit a significant deflection effect on the final-state particles. Therefore, in UPC case, the 2$\left<cos2\phi\right>$ modulation can be solely attributed to the linear polarization of photons.

\begin{figure}
    \centering
    \includegraphics[width=1.0\linewidth]{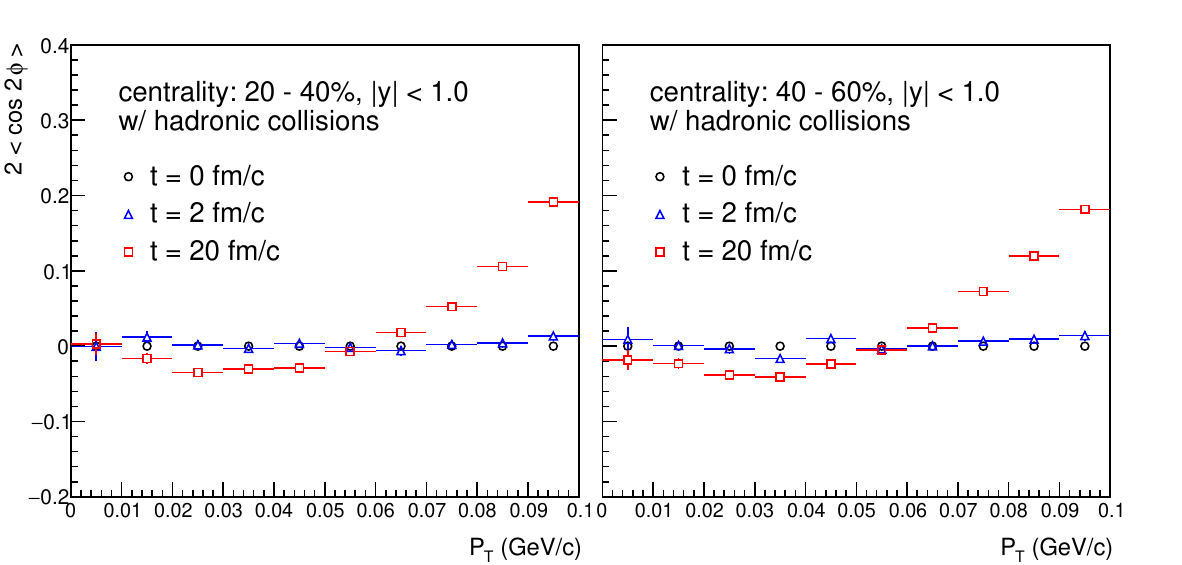}
    \caption{The 2$\left<cos2\phi\right>$ induced by the residual magnetic field in PC case as a function of the $\pi^+\pi^-$ pair $p_T$. The left panel corresponds to centrality 20-40\% class, while the right panel corresponds to centrality 40-60\% class. The black circles represent the results at $t=0$ fm/$c$, the blue triangles at $t=2$ fm/$c$, and the red squares at $t=20$ fm/$c$.}
    \label{PC_cos2phi_pt}
\end{figure}

However, for PC case, the particles generated via strong interactions substantially modify both the spatial configuration of the magnetic field and its temporal decay profile, leading to qualitatively distinct behavior. Figure~\ref{PC_cos2phi_pt} shows the 2$\left<cos2\phi\right>$ distribution as a function of $p_T$ for Au+Au collisions at $\sqrt{s_{\text{NN}}}=200$ GeV with different centrality classes. The black circles show the 2$\left<cos2\phi\right>$ distribution of these $\pi^+ \pi^-$ at $t=0$ fm/$c$, the blue triangles correspond to $t=2$ fm/$c$, and the red squares correspond to $t=20$ fm/$c$. A clear deviation from the baseline (2$\left<cos2\phi\right>=0$) can be observed at the final stage of system evolution. Near $p_T=0.1$ GeV/$c$, the 2$\left<cos2\phi\right>$ induced by the residual magnetic field can even reach 0.2. However, experimentally measured values in UPCs show that the 2$\left<cos2\phi\right>$ modulation induced by polarized photons is only about 0.1 near $p_T=0.1$ GeV/$c$ (due to the finite nuclear radius, this point coincides with the first dip in the $p_T$-dependent 2$\left<cos2\phi\right>$ spectrum). Therefore, the azimuthal modulation originating from the residual magnetic field becomes the dominant contribution. For peripheral collisions, any extraction of nuclear radius information from the $p_T$-dependent 2$\left<cos2\phi\right>$ distribution must rigorously account for distortions induced by this residual magnetic field. We investigated the regime where $t>20$ fm/$c$, beyond which the 2$\left<cos2\phi\right>$ signal remains constant and is therefore not shown here.

\begin{figure}
    \centering
    \includegraphics[width=1.0\linewidth]{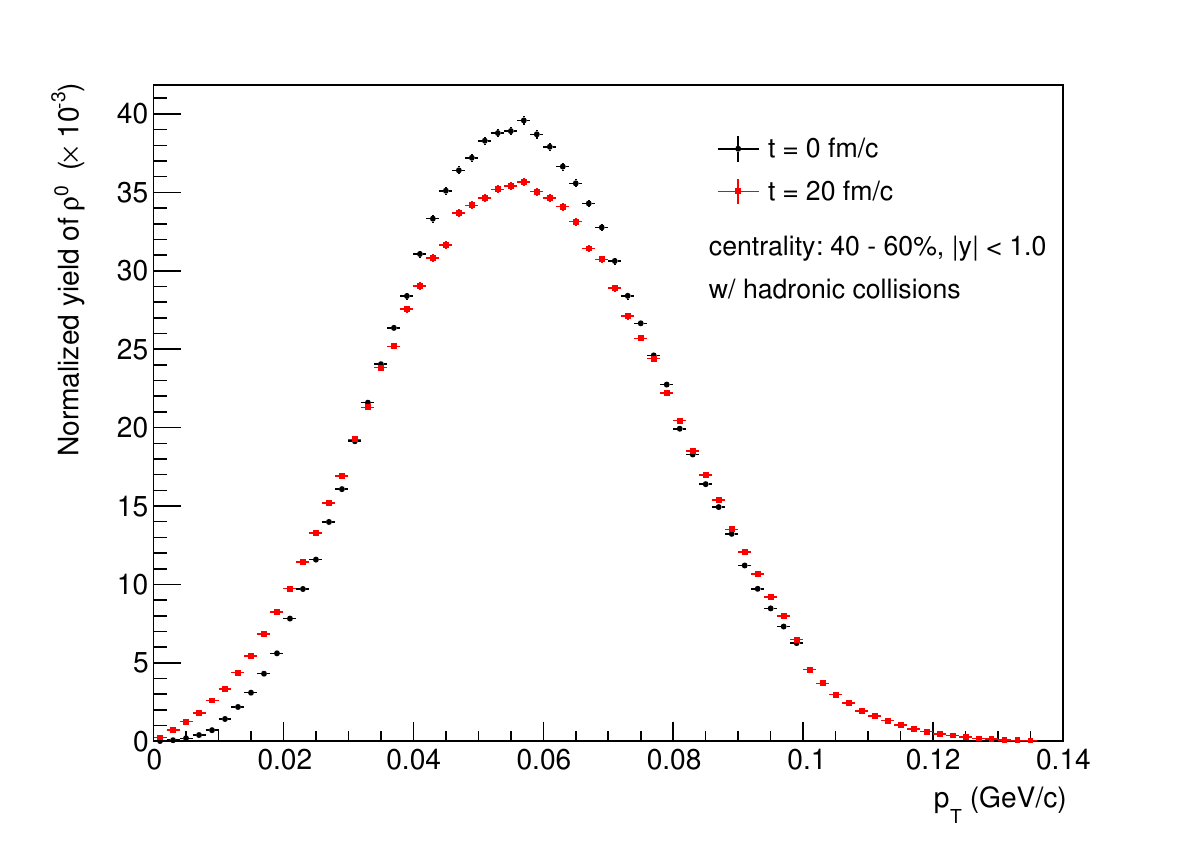}
    \caption{The normalized $p_T$ distribution of the pair in Au+Au collisions at 200 GeV for 40$\sim$60 centrality. Distributions at initial time ($t=0$ fm/$c$) and evolved time ($t=20$ fm/$c$) are displayed as black and red points, respectively.}
    \label{pair_pt}
\end{figure}

In addition to modulating the azimuthal angle, the magnetic field also induces a slight broadening in the pair's $p_T$. Figure~\ref{pair_pt} shows the initial $p_T$ distribution of the pairs (black dots) and their final $p_T$ distribution after system evolution (red dots) in Au+Au collisions at 200 GeV for 40$\sim$60\% centrality. It can be observed that the $p_T$ range broadens from about 0.1 GeV/$c$ to approximately 0.14 GeV/$c$. Since experimental constraints on nuclear shape parameters—derived from measurements of the pair’s $p_T$ distribution~\cite{star2023tomography}—are highly sensitive to the positions of peaks and valleys, this $p_T$ broadening effect may also need to be taken into account.

\section{Summary}
We take the photoproduction of $\rho^0$ mesons in Au+Au collisions at 200 GeV as an example to demonstrate the influence of the residual magnetic field on the azimuthal modulation of final-state particles. For the UPC case, since the magnetic field is concentrated in the central region and decays rapidly, it has no effect on the 2$\left<cos2\phi\right>$ modulation. In contrast, for the PC case, particles produced in the collision are located in the nuclear overlap region and generate an additional magnetic field, altering the spatial distribution of the magnetic field and significantly slowing the decay of the electromagnetic field. In this scenario, when the system evolves to the final state, it causes a noticeable deviation of the 2$\left<cos2\phi\right>$ modulation from the baseline. At $p_T \approx 0.1$ GeV/$c$, the 2$\left<cos2\phi\right>$ modulation induced by the residual magnetic field can reach 0.2, which already exceeds the corresponding 2$\left<cos2\phi\right>$ modulation in UPC measurements caused by the polarization characteristics of photons. Therefore, when extracting nuclear information using 2$\left<cos2\phi\right>$ in PC, the effect of the residual magnetic field must be considered. Moreover, this effect is expected to be widely present in photon-photon collisions and various heavy-ion collision systems (the effect should be more pronounced in Pb-Pb collisions at 5.02 TeV), and thus should be taken into account in related studies.

\section*{Acknowledgments}
This work is supported in part by the National Key Research and Development Program of China under Contract No. 2022YFA1604900 the National Natural Science Foundation of China (NSFC) under Contract No. 12175223 and 12005220. W. Zha is supported by Anhui Provincial Natural Science Foundation No. 2208085J23 and Youth Innovation Promotion Association of Chinese Academy of Sciences (CAS) under Grants No. YSBR-088.

\bibliographystyle{apsrev4-2}
\bibliography{reference}

\begin{thebibliography}{44}%
\makeatletter
\providecommand \@ifxundefined [1]{%
 \@ifx{#1\undefined}
}%
\providecommand \@ifnum [1]{%
 \ifnum #1\expandafter \@firstoftwo
 \else \expandafter \@secondoftwo
 \fi
}%
\providecommand \@ifx [1]{%
 \ifx #1\expandafter \@firstoftwo
 \else \expandafter \@secondoftwo
 \fi
}%
\providecommand \natexlab [1]{#1}%
\providecommand \enquote  [1]{``#1''}%
\providecommand \bibnamefont  [1]{#1}%
\providecommand \bibfnamefont [1]{#1}%
\providecommand \citenamefont [1]{#1}%
\providecommand \href@noop [0]{\@secondoftwo}%
\providecommand \href [0]{\begingroup \@sanitize@url \@href}%
\providecommand \@href[1]{\@@startlink{#1}\@@href}%
\providecommand \@@href[1]{\endgroup#1\@@endlink}%
\providecommand \@sanitize@url [0]{\catcode `\\12\catcode `\$12\catcode `\&12\catcode `\#12\catcode `\^12\catcode `\_12\catcode `\%12\relax}%
\providecommand \@@startlink[1]{}%
\providecommand \@@endlink[0]{}%
\providecommand \url  [0]{\begingroup\@sanitize@url \@url }%
\providecommand \@url [1]{\endgroup\@href {#1}{\urlprefix }}%
\providecommand \urlprefix  [0]{URL }%
\providecommand \Eprint [0]{\href }%
\providecommand \doibase [0]{https://doi.org/}%
\providecommand \selectlanguage [0]{\@gobble}%
\providecommand \bibinfo  [0]{\@secondoftwo}%
\providecommand \bibfield  [0]{\@secondoftwo}%
\providecommand \translation [1]{[#1]}%
\providecommand \BibitemOpen [0]{}%
\providecommand \bibitemStop [0]{}%
\providecommand \bibitemNoStop [0]{.\EOS\space}%
\providecommand \EOS [0]{\spacefactor3000\relax}%
\providecommand \BibitemShut  [1]{\csname bibitem#1\endcsname}%
\let\auto@bib@innerbib\@empty
\bibitem [{\citenamefont {Ma}\ and\ \citenamefont {Zhang}(2023)}]{ma2023influence}%
  \BibitemOpen
  \bibfield  {author} {\bibinfo {author} {\bibfnamefont {Y.}~\bibnamefont {Ma}}\ and\ \bibinfo {author} {\bibfnamefont {S.}~\bibnamefont {Zhang}},\ }in\ \href@noop {} {\emph {\bibinfo {booktitle} {Handbook of Nuclear Physics}}}\ (\bibinfo  {publisher} {Springer},\ \bibinfo {year} {2023})\ pp.\ \bibinfo {pages} {1485--1514}\BibitemShut {NoStop}%
\bibitem [{\citenamefont {Abdallah}\ \emph {et~al.}(2022)\citenamefont {Abdallah}, \citenamefont {Aboona}, \citenamefont {Adam}, \citenamefont {Adamczyk}, \citenamefont {Adams}, \citenamefont {Adkins}, \citenamefont {Agakishiev}, \citenamefont {Aggarwal}, \citenamefont {Aggarwal}, \citenamefont {Ahammed} \emph {et~al.}}]{abdallah2022evidence}%
  \BibitemOpen
  \bibfield  {author} {\bibinfo {author} {\bibfnamefont {M.~S.}\ \bibnamefont {Abdallah}}, \bibinfo {author} {\bibfnamefont {B.}~\bibnamefont {Aboona}}, \bibinfo {author} {\bibfnamefont {J.}~\bibnamefont {Adam}}, \bibinfo {author} {\bibfnamefont {L.}~\bibnamefont {Adamczyk}}, \bibinfo {author} {\bibfnamefont {J.}~\bibnamefont {Adams}}, \bibinfo {author} {\bibfnamefont {J.~K.}\ \bibnamefont {Adkins}}, \bibinfo {author} {\bibfnamefont {G.}~\bibnamefont {Agakishiev}}, \bibinfo {author} {\bibfnamefont {I.}~\bibnamefont {Aggarwal}}, \bibinfo {author} {\bibfnamefont {M.~M.}\ \bibnamefont {Aggarwal}}, \bibinfo {author} {\bibfnamefont {Z.}~\bibnamefont {Ahammed}}, \emph {et~al.},\ }\href@noop {} {\bibfield  {journal} {\bibinfo  {journal} {Physical review letters}\ }\textbf {\bibinfo {volume} {129}},\ \bibinfo {pages} {092501} (\bibinfo {year} {2022})}\BibitemShut {NoStop}%
\bibitem [{\citenamefont {Xu}\ \emph {et~al.}(2018)\citenamefont {Xu}, \citenamefont {Wang}, \citenamefont {Li}, \citenamefont {Zhao}, \citenamefont {Lin}, \citenamefont {Shen},\ and\ \citenamefont {Wang}}]{xu2018importance}%
  \BibitemOpen
  \bibfield  {author} {\bibinfo {author} {\bibfnamefont {H.-J.}\ \bibnamefont {Xu}}, \bibinfo {author} {\bibfnamefont {X.}~\bibnamefont {Wang}}, \bibinfo {author} {\bibfnamefont {H.}~\bibnamefont {Li}}, \bibinfo {author} {\bibfnamefont {J.}~\bibnamefont {Zhao}}, \bibinfo {author} {\bibfnamefont {Z.-W.}\ \bibnamefont {Lin}}, \bibinfo {author} {\bibfnamefont {C.}~\bibnamefont {Shen}},\ and\ \bibinfo {author} {\bibfnamefont {F.}~\bibnamefont {Wang}},\ }\href@noop {} {\bibfield  {journal} {\bibinfo  {journal} {Physical Review Letters}\ }\textbf {\bibinfo {volume} {121}},\ \bibinfo {pages} {022301} (\bibinfo {year} {2018})}\BibitemShut {NoStop}%
\bibitem [{\citenamefont {Lin}\ \emph {et~al.}(2025)\citenamefont {Lin}, \citenamefont {Hu}, \citenamefont {Xu}, \citenamefont {Pu},\ and\ \citenamefont {Wang}}]{lin2025nuclear}%
  \BibitemOpen
  \bibfield  {author} {\bibinfo {author} {\bibfnamefont {S.}~\bibnamefont {Lin}}, \bibinfo {author} {\bibfnamefont {J.}~\bibnamefont {Hu}}, \bibinfo {author} {\bibfnamefont {H.}~\bibnamefont {Xu}}, \bibinfo {author} {\bibfnamefont {S.}~\bibnamefont {Pu}},\ and\ \bibinfo {author} {\bibfnamefont {Q.}~\bibnamefont {Wang}},\ }\href@noop {} {\bibfield  {journal} {\bibinfo  {journal} {Phys. Rev. D}\ }\textbf {\bibinfo {volume} {111}},\ \bibinfo {pages} {074020} (\bibinfo {year} {2025})}\BibitemShut {NoStop}%
\bibitem [{\citenamefont {Collaboration}(2023)}]{star2023tomography}%
  \BibitemOpen
  \bibfield  {author} {\bibinfo {author} {\bibfnamefont {S.}~\bibnamefont {Collaboration}},\ }\href@noop {} {\bibfield  {journal} {\bibinfo  {journal} {Sci. Adv.}\ }\textbf {\bibinfo {volume} {9}},\ \bibinfo {pages} {eabq3903} (\bibinfo {year} {2023})}\BibitemShut {NoStop}%
\bibitem [{\citenamefont {Krauss}\ \emph {et~al.}(1997)\citenamefont {Krauss}, \citenamefont {Greiner},\ and\ \citenamefont {Soff}}]{krauss1997photon}%
  \BibitemOpen
  \bibfield  {author} {\bibinfo {author} {\bibfnamefont {F.}~\bibnamefont {Krauss}}, \bibinfo {author} {\bibfnamefont {M.}~\bibnamefont {Greiner}},\ and\ \bibinfo {author} {\bibfnamefont {G.}~\bibnamefont {Soff}},\ }\href@noop {} {\bibfield  {journal} {\bibinfo  {journal} {Prog. Part. Nucl. Phys.}\ }\textbf {\bibinfo {volume} {39}},\ \bibinfo {pages} {503} (\bibinfo {year} {1997})}\BibitemShut {NoStop}%
\bibitem [{\citenamefont {Bertulani}\ and\ \citenamefont {Baur}(1988)}]{bertulani1988electromagnetic}%
  \BibitemOpen
  \bibfield  {author} {\bibinfo {author} {\bibfnamefont {C.~A.}\ \bibnamefont {Bertulani}}\ and\ \bibinfo {author} {\bibfnamefont {G.}~\bibnamefont {Baur}},\ }\href@noop {} {\bibfield  {journal} {\bibinfo  {journal} {Physics Reports}\ }\textbf {\bibinfo {volume} {163}},\ \bibinfo {pages} {299} (\bibinfo {year} {1988})}\BibitemShut {NoStop}%
\bibitem [{\citenamefont {Criegee}\ \emph {et~al.}(1968)\citenamefont {Criegee}, \citenamefont {Garrell}, \citenamefont {Gottfried}, \citenamefont {Krolzig}, \citenamefont {Loeffler}, \citenamefont {Saulys}, \citenamefont {Sch{\"u}ler}, \citenamefont {Timm}, \citenamefont {Zimmermann}, \citenamefont {Werner} \emph {et~al.}}]{criegee1968rho}%
  \BibitemOpen
  \bibfield  {author} {\bibinfo {author} {\bibfnamefont {L.}~\bibnamefont {Criegee}}, \bibinfo {author} {\bibfnamefont {M.~H.}\ \bibnamefont {Garrell}}, \bibinfo {author} {\bibfnamefont {C.}~\bibnamefont {Gottfried}}, \bibinfo {author} {\bibfnamefont {A.}~\bibnamefont {Krolzig}}, \bibinfo {author} {\bibfnamefont {G.}~\bibnamefont {Loeffler}}, \bibinfo {author} {\bibfnamefont {A.}~\bibnamefont {Saulys}}, \bibinfo {author} {\bibfnamefont {K.~P.}\ \bibnamefont {Sch{\"u}ler}}, \bibinfo {author} {\bibfnamefont {U.}~\bibnamefont {Timm}}, \bibinfo {author} {\bibfnamefont {W.}~\bibnamefont {Zimmermann}}, \bibinfo {author} {\bibfnamefont {H.}~\bibnamefont {Werner}}, \emph {et~al.},\ }\href@noop {} {\bibfield  {journal} {\bibinfo  {journal} {Phys. Lett. B}\ }\textbf {\bibinfo {volume} {28}},\ \bibinfo {pages} {282} (\bibinfo {year} {1968})}\BibitemShut {NoStop}%
\bibitem [{\citenamefont {Ballam}\ \emph {et~al.}(1970)\citenamefont {Ballam}, \citenamefont {Chadwick}, \citenamefont {Gearhart}, \citenamefont {Guiragossi\'an}, \citenamefont {Menke}, \citenamefont {Murray}, \citenamefont {Seyboth}, \citenamefont {Shapira}, \citenamefont {Sinclair}, \citenamefont {Skillicorn}, \citenamefont {Wolf}, \citenamefont {Milburn}, \citenamefont {Bingham}, \citenamefont {Fretter}, \citenamefont {Moffeit}, \citenamefont {Podolsky}, \citenamefont {Rabin}, \citenamefont {Rosenfeld},\ and\ \citenamefont {Windmolders}}]{PhysRevLett.24.1364}%
  \BibitemOpen
  \bibfield  {author} {\bibinfo {author} {\bibfnamefont {J.}~\bibnamefont {Ballam}}, \bibinfo {author} {\bibfnamefont {G.~B.}\ \bibnamefont {Chadwick}}, \bibinfo {author} {\bibfnamefont {R.}~\bibnamefont {Gearhart}}, \bibinfo {author} {\bibfnamefont {Z.~G.~T.}\ \bibnamefont {Guiragossi\'an}}, \bibinfo {author} {\bibfnamefont {M.}~\bibnamefont {Menke}}, \bibinfo {author} {\bibfnamefont {J.~J.}\ \bibnamefont {Murray}}, \bibinfo {author} {\bibfnamefont {P.}~\bibnamefont {Seyboth}}, \bibinfo {author} {\bibfnamefont {A.}~\bibnamefont {Shapira}}, \bibinfo {author} {\bibfnamefont {C.~K.}\ \bibnamefont {Sinclair}}, \bibinfo {author} {\bibfnamefont {I.~O.}\ \bibnamefont {Skillicorn}}, \bibinfo {author} {\bibfnamefont {G.}~\bibnamefont {Wolf}}, \bibinfo {author} {\bibfnamefont {R.~H.}\ \bibnamefont {Milburn}}, \bibinfo {author} {\bibfnamefont {H.~H.}\ \bibnamefont {Bingham}}, \bibinfo {author} {\bibfnamefont {W.~B.}\ \bibnamefont {Fretter}}, \bibinfo {author} {\bibfnamefont {K.~C.}\ \bibnamefont {Moffeit}}, \bibinfo
  {author} {\bibfnamefont {W.~J.}\ \bibnamefont {Podolsky}}, \bibinfo {author} {\bibfnamefont {M.~S.}\ \bibnamefont {Rabin}}, \bibinfo {author} {\bibfnamefont {A.~H.}\ \bibnamefont {Rosenfeld}},\ and\ \bibinfo {author} {\bibfnamefont {R.}~\bibnamefont {Windmolders}},\ }\href {https://doi.org/10.1103/PhysRevLett.24.1364} {\bibfield  {journal} {\bibinfo  {journal} {Phys. Rev. Lett.}\ }\textbf {\bibinfo {volume} {24}},\ \bibinfo {pages} {1364} (\bibinfo {year} {1970})}\BibitemShut {NoStop}%
\bibitem [{\citenamefont {Eisenberg}\ \emph {et~al.}(1976)\citenamefont {Eisenberg}, \citenamefont {Haber}, \citenamefont {Kogan}, \citenamefont {Karshon}, \citenamefont {Ronat}, \citenamefont {Shapira},\ and\ \citenamefont {Yekutieli}}]{Eisenberg:1975jr}%
  \BibitemOpen
  \bibfield  {author} {\bibinfo {author} {\bibfnamefont {Y.}~\bibnamefont {Eisenberg}}, \bibinfo {author} {\bibfnamefont {B.}~\bibnamefont {Haber}}, \bibinfo {author} {\bibfnamefont {E.}~\bibnamefont {Kogan}}, \bibinfo {author} {\bibfnamefont {U.}~\bibnamefont {Karshon}}, \bibinfo {author} {\bibfnamefont {E.~E.}\ \bibnamefont {Ronat}}, \bibinfo {author} {\bibfnamefont {A.}~\bibnamefont {Shapira}},\ and\ \bibinfo {author} {\bibfnamefont {G.}~\bibnamefont {Yekutieli}},\ }\href {https://doi.org/10.1016/0550-3213(76)90073-0} {\bibfield  {journal} {\bibinfo  {journal} {Nucl. Phys. B}\ }\textbf {\bibinfo {volume} {104}},\ \bibinfo {pages} {61} (\bibinfo {year} {1976})}\BibitemShut {NoStop}%
\bibitem [{\citenamefont {Derrick}\ \emph {et~al.}(1996)\citenamefont {Derrick} \emph {et~al.}}]{ZEUS:1996esk}%
  \BibitemOpen
  \bibfield  {author} {\bibinfo {author} {\bibfnamefont {M.}~\bibnamefont {Derrick}} \emph {et~al.} (\bibinfo {collaboration} {ZEUS}),\ }\href {https://doi.org/10.1016/0370-2693(96)00172-4} {\bibfield  {journal} {\bibinfo  {journal} {Phys. Lett. B}\ }\textbf {\bibinfo {volume} {377}},\ \bibinfo {pages} {259} (\bibinfo {year} {1996})},\ \Eprint {https://arxiv.org/abs/hep-ex/9601009} {arXiv:hep-ex/9601009} \BibitemShut {NoStop}%
\bibitem [{\citenamefont {Abelev}\ \emph {et~al.}(2008)\citenamefont {Abelev} \emph {et~al.}}]{STAR:2007elq}%
  \BibitemOpen
  \bibfield  {author} {\bibinfo {author} {\bibfnamefont {B.~I.}\ \bibnamefont {Abelev}} \emph {et~al.} (\bibinfo {collaboration} {STAR}),\ }\href {https://doi.org/10.1103/PhysRevC.77.034910} {\bibfield  {journal} {\bibinfo  {journal} {Phys. Rev. C}\ }\textbf {\bibinfo {volume} {77}},\ \bibinfo {pages} {034910} (\bibinfo {year} {2008})},\ \Eprint {https://arxiv.org/abs/0712.3320} {arXiv:0712.3320 [nucl-ex]} \BibitemShut {NoStop}%
\bibitem [{\citenamefont {Hiraiwa}\ \emph {et~al.}(2018)\citenamefont {Hiraiwa} \emph {et~al.}}]{LEPS:2017nqz}%
  \BibitemOpen
  \bibfield  {author} {\bibinfo {author} {\bibfnamefont {T.}~\bibnamefont {Hiraiwa}} \emph {et~al.} (\bibinfo {collaboration} {LEPS}),\ }\href {https://doi.org/10.1103/PhysRevC.97.035208} {\bibfield  {journal} {\bibinfo  {journal} {Phys. Rev. C}\ }\textbf {\bibinfo {volume} {97}},\ \bibinfo {pages} {035208} (\bibinfo {year} {2018})},\ \Eprint {https://arxiv.org/abs/1711.01095} {arXiv:1711.01095 [nucl-ex]} \BibitemShut {NoStop}%
\bibitem [{\citenamefont {Zha}\ \emph {et~al.}(2021)\citenamefont {Zha}, \citenamefont {Brandenburg}, \citenamefont {Ruan},\ and\ \citenamefont {Tang}}]{zha2021exploring}%
  \BibitemOpen
  \bibfield  {author} {\bibinfo {author} {\bibfnamefont {W.}~\bibnamefont {Zha}}, \bibinfo {author} {\bibfnamefont {J.~D.}\ \bibnamefont {Brandenburg}}, \bibinfo {author} {\bibfnamefont {L.}~\bibnamefont {Ruan}},\ and\ \bibinfo {author} {\bibfnamefont {Z.}~\bibnamefont {Tang}},\ }\href@noop {} {\bibfield  {journal} {\bibinfo  {journal} {Phys. Rev. D}\ }\textbf {\bibinfo {volume} {103}},\ \bibinfo {pages} {033007} (\bibinfo {year} {2021})}\BibitemShut {NoStop}%
\bibitem [{\citenamefont {Xing}\ \emph {et~al.}(2020{\natexlab{a}})\citenamefont {Xing}, \citenamefont {Zhang}, \citenamefont {Zhou},\ and\ \citenamefont {Zhou}}]{Xing:2020hwh}%
  \BibitemOpen
  \bibfield  {author} {\bibinfo {author} {\bibfnamefont {H.}~\bibnamefont {Xing}}, \bibinfo {author} {\bibfnamefont {C.}~\bibnamefont {Zhang}}, \bibinfo {author} {\bibfnamefont {J.}~\bibnamefont {Zhou}},\ and\ \bibinfo {author} {\bibfnamefont {Y.-J.}\ \bibnamefont {Zhou}},\ }\href {https://doi.org/10.1007/JHEP10(2020)064} {\bibfield  {journal} {\bibinfo  {journal} {JHEP}\ }\textbf {\bibinfo {volume} {10}},\ \bibinfo {pages} {064}},\ \Eprint {https://arxiv.org/abs/2006.06206} {arXiv:2006.06206 [hep-ph]} \BibitemShut {NoStop}%
\bibitem [{\citenamefont {Xing}\ \emph {et~al.}(2020{\natexlab{b}})\citenamefont {Xing}, \citenamefont {Zhang}, \citenamefont {Zhou},\ and\ \citenamefont {Zhou}}]{xing2020cos}%
  \BibitemOpen
  \bibfield  {author} {\bibinfo {author} {\bibfnamefont {H.}~\bibnamefont {Xing}}, \bibinfo {author} {\bibfnamefont {C.}~\bibnamefont {Zhang}}, \bibinfo {author} {\bibfnamefont {J.}~\bibnamefont {Zhou}},\ and\ \bibinfo {author} {\bibfnamefont {Y.}~\bibnamefont {Zhou}},\ }\href@noop {} {\bibfield  {journal} {\bibinfo  {journal} {J. High Energy Phys.}\ }\textbf {\bibinfo {volume} {64}}\bibinfo  {number} { (10)},\ \bibinfo {pages} {1}}\BibitemShut {NoStop}%
\bibitem [{\citenamefont {Collaboration}\ \emph {et~al.}(2024)\citenamefont {Collaboration} \emph {et~al.}}]{alice2024measurement}%
  \BibitemOpen
\bibfield  {number} {  }\bibfield  {author} {\bibinfo {author} {\bibfnamefont {A.}~\bibnamefont {Collaboration}} \emph {et~al.},\ }\href@noop {} {\bibfield  {journal} {\bibinfo  {journal} {Phys. Lett. B}\ }\textbf {\bibinfo {volume} {858}},\ \bibinfo {pages} {139017} (\bibinfo {year} {2024})}\BibitemShut {NoStop}%
\bibitem [{\citenamefont {Wu}\ \emph {et~al.}(2022)\citenamefont {Wu}, \citenamefont {Li}, \citenamefont {Tang}, \citenamefont {Wang},\ and\ \citenamefont {Zha}}]{wu2022reaction}%
  \BibitemOpen
  \bibfield  {author} {\bibinfo {author} {\bibfnamefont {X.}~\bibnamefont {Wu}}, \bibinfo {author} {\bibfnamefont {X.}~\bibnamefont {Li}}, \bibinfo {author} {\bibfnamefont {Z.}~\bibnamefont {Tang}}, \bibinfo {author} {\bibfnamefont {P.}~\bibnamefont {Wang}},\ and\ \bibinfo {author} {\bibfnamefont {W.}~\bibnamefont {Zha}},\ }\href@noop {} {\bibfield  {journal} {\bibinfo  {journal} {Phys. Rev. Res.}\ }\textbf {\bibinfo {volume} {4}},\ \bibinfo {pages} {L042048} (\bibinfo {year} {2022})}\BibitemShut {NoStop}%
\bibitem [{\citenamefont {Kharzeev}\ \emph {et~al.}(2008)\citenamefont {Kharzeev}, \citenamefont {McLerran},\ and\ \citenamefont {Warringa}}]{kharzeev2008effects}%
  \BibitemOpen
  \bibfield  {author} {\bibinfo {author} {\bibfnamefont {D.~E.}\ \bibnamefont {Kharzeev}}, \bibinfo {author} {\bibfnamefont {L.~D.}\ \bibnamefont {McLerran}},\ and\ \bibinfo {author} {\bibfnamefont {H.~J.}\ \bibnamefont {Warringa}},\ }\href@noop {} {\bibfield  {journal} {\bibinfo  {journal} {Nuclear Physics A}\ }\textbf {\bibinfo {volume} {803}},\ \bibinfo {pages} {227} (\bibinfo {year} {2008})}\BibitemShut {NoStop}%
\bibitem [{\citenamefont {Deng}\ and\ \citenamefont {Huang}(2012)}]{deng2012event}%
  \BibitemOpen
  \bibfield  {author} {\bibinfo {author} {\bibfnamefont {W.-T.}\ \bibnamefont {Deng}}\ and\ \bibinfo {author} {\bibfnamefont {X.-G.}\ \bibnamefont {Huang}},\ }\href@noop {} {\bibfield  {journal} {\bibinfo  {journal} {Physical Review C—Nuclear Physics}\ }\textbf {\bibinfo {volume} {85}},\ \bibinfo {pages} {044907} (\bibinfo {year} {2012})}\BibitemShut {NoStop}%
\bibitem [{\citenamefont {Toneev}\ \emph {et~al.}(2012)\citenamefont {Toneev}, \citenamefont {Konchakovski}, \citenamefont {Voronyuk}, \citenamefont {Bratkovskaya},\ and\ \citenamefont {Cassing}}]{Toneev:2012zx}%
  \BibitemOpen
  \bibfield  {author} {\bibinfo {author} {\bibfnamefont {V.~D.}\ \bibnamefont {Toneev}}, \bibinfo {author} {\bibfnamefont {V.~P.}\ \bibnamefont {Konchakovski}}, \bibinfo {author} {\bibfnamefont {V.}~\bibnamefont {Voronyuk}}, \bibinfo {author} {\bibfnamefont {E.~L.}\ \bibnamefont {Bratkovskaya}},\ and\ \bibinfo {author} {\bibfnamefont {W.}~\bibnamefont {Cassing}},\ }\href {https://doi.org/10.1103/PhysRevC.86.064907} {\bibfield  {journal} {\bibinfo  {journal} {Phys. Rev. C}\ }\textbf {\bibinfo {volume} {86}},\ \bibinfo {pages} {064907} (\bibinfo {year} {2012})},\ \Eprint {https://arxiv.org/abs/1208.2519} {arXiv:1208.2519 [nucl-th]} \BibitemShut {NoStop}%
\bibitem [{\citenamefont {Chen}\ \emph {et~al.}(2024)\citenamefont {Chen} \emph {et~al.}}]{Chen:2024aom}%
  \BibitemOpen
  \bibfield  {author} {\bibinfo {author} {\bibfnamefont {J.}~\bibnamefont {Chen}} \emph {et~al.},\ }\href {https://doi.org/10.1007/s41365-024-01591-2} {\bibfield  {journal} {\bibinfo  {journal} {Nucl. Sci. Tech.}\ }\textbf {\bibinfo {volume} {35}},\ \bibinfo {pages} {214} (\bibinfo {year} {2024})},\ \Eprint {https://arxiv.org/abs/2407.02935} {arXiv:2407.02935 [nucl-ex]} \BibitemShut {NoStop}%
\bibitem [{\citenamefont {Bass}\ \emph {et~al.}(1998)\citenamefont {Bass}, \citenamefont {Belkacem}, \citenamefont {Bleicher}, \citenamefont {Brandstetter}, \citenamefont {Bravina}, \citenamefont {Ernst}, \citenamefont {Gerland}, \citenamefont {Hofmann}, \citenamefont {Hofmann}, \citenamefont {Konopka} \emph {et~al.}}]{bass1998microscopic}%
  \BibitemOpen
  \bibfield  {author} {\bibinfo {author} {\bibfnamefont {S.~A.}\ \bibnamefont {Bass}}, \bibinfo {author} {\bibfnamefont {M.}~\bibnamefont {Belkacem}}, \bibinfo {author} {\bibfnamefont {M.}~\bibnamefont {Bleicher}}, \bibinfo {author} {\bibfnamefont {M.}~\bibnamefont {Brandstetter}}, \bibinfo {author} {\bibfnamefont {L.}~\bibnamefont {Bravina}}, \bibinfo {author} {\bibfnamefont {C.}~\bibnamefont {Ernst}}, \bibinfo {author} {\bibfnamefont {L.}~\bibnamefont {Gerland}}, \bibinfo {author} {\bibfnamefont {M.}~\bibnamefont {Hofmann}}, \bibinfo {author} {\bibfnamefont {S.}~\bibnamefont {Hofmann}}, \bibinfo {author} {\bibfnamefont {J.}~\bibnamefont {Konopka}}, \emph {et~al.},\ }\href@noop {} {\bibfield  {journal} {\bibinfo  {journal} {Prog. Part. Nucl. Phys.}\ }\textbf {\bibinfo {volume} {41}},\ \bibinfo {pages} {255} (\bibinfo {year} {1998})}\BibitemShut {NoStop}%
\bibitem [{\citenamefont {Bleicher}\ \emph {et~al.}(1999)\citenamefont {Bleicher}, \citenamefont {Zabrodin}, \citenamefont {Spieles}, \citenamefont {Bass}, \citenamefont {Ernst}, \citenamefont {Soff}, \citenamefont {Bravina}, \citenamefont {Belkacem}, \citenamefont {Weber}, \citenamefont {St{\"o}cker} \emph {et~al.}}]{bleicher1999relativistic}%
  \BibitemOpen
  \bibfield  {author} {\bibinfo {author} {\bibfnamefont {M.}~\bibnamefont {Bleicher}}, \bibinfo {author} {\bibfnamefont {E.}~\bibnamefont {Zabrodin}}, \bibinfo {author} {\bibfnamefont {C.}~\bibnamefont {Spieles}}, \bibinfo {author} {\bibfnamefont {S.~A.}\ \bibnamefont {Bass}}, \bibinfo {author} {\bibfnamefont {C.}~\bibnamefont {Ernst}}, \bibinfo {author} {\bibfnamefont {S.}~\bibnamefont {Soff}}, \bibinfo {author} {\bibfnamefont {L.}~\bibnamefont {Bravina}}, \bibinfo {author} {\bibfnamefont {M.}~\bibnamefont {Belkacem}}, \bibinfo {author} {\bibfnamefont {H.}~\bibnamefont {Weber}}, \bibinfo {author} {\bibfnamefont {H.}~\bibnamefont {St{\"o}cker}}, \emph {et~al.},\ }\href@noop {} {\bibfield  {journal} {\bibinfo  {journal} {Journal of Physics G: Nuclear and Particle Physics}\ }\textbf {\bibinfo {volume} {25}},\ \bibinfo {pages} {1859} (\bibinfo {year} {1999})}\BibitemShut {NoStop}%
\bibitem [{\citenamefont {Li}\ \emph {et~al.}(2020)\citenamefont {Li}, \citenamefont {Zhou},\ and\ \citenamefont {Zhou}}]{Li:2019sin}%
  \BibitemOpen
  \bibfield  {author} {\bibinfo {author} {\bibfnamefont {C.}~\bibnamefont {Li}}, \bibinfo {author} {\bibfnamefont {J.}~\bibnamefont {Zhou}},\ and\ \bibinfo {author} {\bibfnamefont {Y.-J.}\ \bibnamefont {Zhou}},\ }\href {https://doi.org/10.1103/PhysRevD.101.034015} {\bibfield  {journal} {\bibinfo  {journal} {Phys. Rev. D}\ }\textbf {\bibinfo {volume} {101}},\ \bibinfo {pages} {034015} (\bibinfo {year} {2020})},\ \Eprint {https://arxiv.org/abs/1911.00237} {arXiv:1911.00237 [hep-ph]} \BibitemShut {NoStop}%
\bibitem [{\citenamefont {Klein}\ \emph {et~al.}(2019)\citenamefont {Klein}, \citenamefont {Mueller}, \citenamefont {Xiao},\ and\ \citenamefont {Yuan}}]{Klein:2018fmp}%
  \BibitemOpen
  \bibfield  {author} {\bibinfo {author} {\bibfnamefont {S.}~\bibnamefont {Klein}}, \bibinfo {author} {\bibfnamefont {A.~H.}\ \bibnamefont {Mueller}}, \bibinfo {author} {\bibfnamefont {B.-W.}\ \bibnamefont {Xiao}},\ and\ \bibinfo {author} {\bibfnamefont {F.}~\bibnamefont {Yuan}},\ }\href {https://doi.org/10.1103/PhysRevLett.122.132301} {\bibfield  {journal} {\bibinfo  {journal} {Phys. Rev. Lett.}\ }\textbf {\bibinfo {volume} {122}},\ \bibinfo {pages} {132301} (\bibinfo {year} {2019})},\ \Eprint {https://arxiv.org/abs/1811.05519} {arXiv:1811.05519 [hep-ph]} \BibitemShut {NoStop}%
\bibitem [{\citenamefont {Von~Weizsacker}(1934)}]{von1934radiation}%
  \BibitemOpen
  \bibfield  {author} {\bibinfo {author} {\bibfnamefont {C.}~\bibnamefont {Von~Weizsacker}},\ }\href@noop {} {\bibfield  {journal} {\bibinfo  {journal} {Z. Phys}\ }\textbf {\bibinfo {volume} {88}},\ \bibinfo {pages} {95} (\bibinfo {year} {1934})}\BibitemShut {NoStop}%
\bibitem [{\citenamefont {Williams}(1934)}]{williams1934nature}%
  \BibitemOpen
  \bibfield  {author} {\bibinfo {author} {\bibfnamefont {E.}~\bibnamefont {Williams}},\ }\href@noop {} {\bibfield  {journal} {\bibinfo  {journal} {Physical Review}\ }\textbf {\bibinfo {volume} {45}},\ \bibinfo {pages} {729} (\bibinfo {year} {1934})}\BibitemShut {NoStop}%
\bibitem [{\citenamefont {Barrett}\ and\ \citenamefont {Jackson}(1977)}]{barrett1977nuclear}%
  \BibitemOpen
  \bibfield  {author} {\bibinfo {author} {\bibfnamefont {R.~C.}\ \bibnamefont {Barrett}}\ and\ \bibinfo {author} {\bibfnamefont {D.~F.}\ \bibnamefont {Jackson}},\ }\href@noop {} {\emph {\bibinfo {title} {Nuclear Sizes and Structure}}}\ (\bibinfo  {publisher} {Clarendon Press},\ \bibinfo {address} {Oxford},\ \bibinfo {year} {1977})\BibitemShut {NoStop}%
\bibitem [{\citenamefont {Miller}\ \emph {et~al.}(2007)\citenamefont {Miller}, \citenamefont {Reygers}, \citenamefont {Sanders},\ and\ \citenamefont {Steinberg}}]{miller2007glauber}%
  \BibitemOpen
  \bibfield  {author} {\bibinfo {author} {\bibfnamefont {M.~L.}\ \bibnamefont {Miller}}, \bibinfo {author} {\bibfnamefont {K.}~\bibnamefont {Reygers}}, \bibinfo {author} {\bibfnamefont {S.~J.}\ \bibnamefont {Sanders}},\ and\ \bibinfo {author} {\bibfnamefont {P.}~\bibnamefont {Steinberg}},\ }\href@noop {} {\bibfield  {journal} {\bibinfo  {journal} {Annu. Rev. Nucl. Part. Sci.}\ }\textbf {\bibinfo {volume} {57}},\ \bibinfo {pages} {205} (\bibinfo {year} {2007})}\BibitemShut {NoStop}%
\bibitem [{\citenamefont {Bauer}\ \emph {et~al.}(1978)\citenamefont {Bauer}, \citenamefont {Spital}, \citenamefont {Yennie},\ and\ \citenamefont {Pipkin}}]{bauer1978hadronic}%
  \BibitemOpen
  \bibfield  {author} {\bibinfo {author} {\bibfnamefont {T.~H.}\ \bibnamefont {Bauer}}, \bibinfo {author} {\bibfnamefont {R.~D.}\ \bibnamefont {Spital}}, \bibinfo {author} {\bibfnamefont {D.~R.}\ \bibnamefont {Yennie}},\ and\ \bibinfo {author} {\bibfnamefont {F.~M.}\ \bibnamefont {Pipkin}},\ }\href@noop {} {\bibfield  {journal} {\bibinfo  {journal} {Rev. Mod. Phys.}\ }\textbf {\bibinfo {volume} {50}},\ \bibinfo {pages} {261} (\bibinfo {year} {1978})}\BibitemShut {NoStop}%
\bibitem [{\citenamefont {Klein}\ and\ \citenamefont {Nystrand}(1999)}]{Klein:1999qj}%
  \BibitemOpen
  \bibfield  {author} {\bibinfo {author} {\bibfnamefont {S.}~\bibnamefont {Klein}}\ and\ \bibinfo {author} {\bibfnamefont {J.}~\bibnamefont {Nystrand}},\ }\href {https://doi.org/10.1103/PhysRevC.60.014903} {\bibfield  {journal} {\bibinfo  {journal} {Phys. Rev. C}\ }\textbf {\bibinfo {volume} {60}},\ \bibinfo {pages} {014903} (\bibinfo {year} {1999})},\ \Eprint {https://arxiv.org/abs/hep-ph/9902259} {arXiv:hep-ph/9902259} \BibitemShut {NoStop}%
\bibitem [{\citenamefont {Hufner}\ and\ \citenamefont {Kopeliovich}(1998)}]{Hufner:1997jg}%
  \BibitemOpen
  \bibfield  {author} {\bibinfo {author} {\bibfnamefont {J.}~\bibnamefont {Hufner}}\ and\ \bibinfo {author} {\bibfnamefont {B.~Z.}\ \bibnamefont {Kopeliovich}},\ }\href {https://doi.org/10.1016/S0370-2693(98)00257-3} {\bibfield  {journal} {\bibinfo  {journal} {Phys. Lett. B}\ }\textbf {\bibinfo {volume} {426}},\ \bibinfo {pages} {154} (\bibinfo {year} {1998})},\ \Eprint {https://arxiv.org/abs/hep-ph/9712297} {arXiv:hep-ph/9712297} \BibitemShut {NoStop}%
\bibitem [{\citenamefont {Zha}\ \emph {et~al.}(2018)\citenamefont {Zha}, \citenamefont {Klein}, \citenamefont {Ma}, \citenamefont {Ruan}, \citenamefont {Todoroki}, \citenamefont {Tang}, \citenamefont {Xu}, \citenamefont {Yang}, \citenamefont {Yang},\ and\ \citenamefont {Yang}}]{zha2018coherent}%
  \BibitemOpen
  \bibfield  {author} {\bibinfo {author} {\bibfnamefont {W.}~\bibnamefont {Zha}}, \bibinfo {author} {\bibfnamefont {S.~R.}\ \bibnamefont {Klein}}, \bibinfo {author} {\bibfnamefont {R.}~\bibnamefont {Ma}}, \bibinfo {author} {\bibfnamefont {L.}~\bibnamefont {Ruan}}, \bibinfo {author} {\bibfnamefont {T.}~\bibnamefont {Todoroki}}, \bibinfo {author} {\bibfnamefont {Z.}~\bibnamefont {Tang}}, \bibinfo {author} {\bibfnamefont {Z.}~\bibnamefont {Xu}}, \bibinfo {author} {\bibfnamefont {C.}~\bibnamefont {Yang}}, \bibinfo {author} {\bibfnamefont {Q.}~\bibnamefont {Yang}},\ and\ \bibinfo {author} {\bibfnamefont {S.}~\bibnamefont {Yang}},\ }\href@noop {} {\bibfield  {journal} {\bibinfo  {journal} {Phys. Rev. C}\ }\textbf {\bibinfo {volume} {97}},\ \bibinfo {pages} {044910} (\bibinfo {year} {2018})}\BibitemShut {NoStop}%
\bibitem [{\citenamefont {Veyssiere}\ \emph {et~al.}(1970)\citenamefont {Veyssiere}, \citenamefont {Beil}, \citenamefont {Bergère}, \citenamefont {Carlos},\ and\ \citenamefont {Lepretre}}]{veyssiere1970photoneutron}%
  \BibitemOpen
  \bibfield  {author} {\bibinfo {author} {\bibfnamefont {A.}~\bibnamefont {Veyssiere}}, \bibinfo {author} {\bibfnamefont {H.}~\bibnamefont {Beil}}, \bibinfo {author} {\bibfnamefont {R.}~\bibnamefont {Bergère}}, \bibinfo {author} {\bibfnamefont {P.}~\bibnamefont {Carlos}},\ and\ \bibinfo {author} {\bibfnamefont {A.}~\bibnamefont {Lepretre}},\ }\href@noop {} {\bibfield  {journal} {\bibinfo  {journal} {Nucl. Phys. A}\ }\textbf {\bibinfo {volume} {159}},\ \bibinfo {pages} {561} (\bibinfo {year} {1970})}\BibitemShut {NoStop}%
\bibitem [{\citenamefont {Lepretre}\ \emph {et~al.}(1981)\citenamefont {Lepretre}, \citenamefont {Beil}, \citenamefont {Bergère}, \citenamefont {Carlos}, \citenamefont {Fagot}, \citenamefont {Miniac},\ and\ \citenamefont {Veyssiere}}]{lepretre1981measurements}%
  \BibitemOpen
  \bibfield  {author} {\bibinfo {author} {\bibfnamefont {A.}~\bibnamefont {Lepretre}}, \bibinfo {author} {\bibfnamefont {H.}~\bibnamefont {Beil}}, \bibinfo {author} {\bibfnamefont {R.}~\bibnamefont {Bergère}}, \bibinfo {author} {\bibfnamefont {P.}~\bibnamefont {Carlos}}, \bibinfo {author} {\bibfnamefont {J.}~\bibnamefont {Fagot}}, \bibinfo {author} {\bibfnamefont {A.~D.}\ \bibnamefont {Miniac}},\ and\ \bibinfo {author} {\bibfnamefont {A.}~\bibnamefont {Veyssiere}},\ }\href@noop {} {\bibfield  {journal} {\bibinfo  {journal} {Nucl. Phys. A}\ }\textbf {\bibinfo {volume} {367}},\ \bibinfo {pages} {237} (\bibinfo {year} {1981})}\BibitemShut {NoStop}%
\bibitem [{\citenamefont {Carlos}\ \emph {et~al.}(1984)\citenamefont {Carlos}, \citenamefont {Beil}, \citenamefont {Bergère}, \citenamefont {Fagot}, \citenamefont {Lepretre},\ and\ \citenamefont {Veyssière}}]{carlos1984total}%
  \BibitemOpen
  \bibfield  {author} {\bibinfo {author} {\bibfnamefont {P.}~\bibnamefont {Carlos}}, \bibinfo {author} {\bibfnamefont {H.}~\bibnamefont {Beil}}, \bibinfo {author} {\bibfnamefont {R.}~\bibnamefont {Bergère}}, \bibinfo {author} {\bibfnamefont {J.}~\bibnamefont {Fagot}}, \bibinfo {author} {\bibfnamefont {A.}~\bibnamefont {Lepretre}},\ and\ \bibinfo {author} {\bibfnamefont {A.}~\bibnamefont {Veyssière}},\ }\href@noop {} {\bibfield  {journal} {\bibinfo  {journal} {Nucl. Phys. A}\ }\textbf {\bibinfo {volume} {431}},\ \bibinfo {pages} {573} (\bibinfo {year} {1984})}\BibitemShut {NoStop}%
\bibitem [{\citenamefont {Armstrong}\ \emph {et~al.}(1972{\natexlab{a}})\citenamefont {Armstrong}, \citenamefont {Hogg}, \citenamefont {Lewis}, \citenamefont {Robertson}, \citenamefont {Brookes}, \citenamefont {Clough}, \citenamefont {Freeland}, \citenamefont {Galbraith}, \citenamefont {King},\ and\ \citenamefont {Rawlinson}}]{armstrong1972total}%
  \BibitemOpen
  \bibfield  {author} {\bibinfo {author} {\bibfnamefont {T.~A.}\ \bibnamefont {Armstrong}}, \bibinfo {author} {\bibfnamefont {W.~R.}\ \bibnamefont {Hogg}}, \bibinfo {author} {\bibfnamefont {G.~M.}\ \bibnamefont {Lewis}}, \bibinfo {author} {\bibfnamefont {A.~W.}\ \bibnamefont {Robertson}}, \bibinfo {author} {\bibfnamefont {G.~R.}\ \bibnamefont {Brookes}}, \bibinfo {author} {\bibfnamefont {A.~S.}\ \bibnamefont {Clough}}, \bibinfo {author} {\bibfnamefont {J.~H.}\ \bibnamefont {Freeland}}, \bibinfo {author} {\bibfnamefont {W.}~\bibnamefont {Galbraith}}, \bibinfo {author} {\bibfnamefont {A.~F.}\ \bibnamefont {King}},\ and\ \bibinfo {author} {\bibfnamefont {W.~R.}\ \bibnamefont {Rawlinson}},\ }\href@noop {} {\bibfield  {journal} {\bibinfo  {journal} {Phys. Rev. D}\ }\textbf {\bibinfo {volume} {5}},\ \bibinfo {pages} {1640} (\bibinfo {year} {1972}{\natexlab{a}})}\BibitemShut {NoStop}%
\bibitem [{\citenamefont {Caldwell}\ \emph {et~al.}(1973)\citenamefont {Caldwell}, \citenamefont {Elings}, \citenamefont {Hesse}, \citenamefont {Morrison}, \citenamefont {Murphy},\ and\ \citenamefont {Yount}}]{caldwell1973total}%
  \BibitemOpen
  \bibfield  {author} {\bibinfo {author} {\bibfnamefont {D.~O.}\ \bibnamefont {Caldwell}}, \bibinfo {author} {\bibfnamefont {V.~B.}\ \bibnamefont {Elings}}, \bibinfo {author} {\bibfnamefont {W.~P.}\ \bibnamefont {Hesse}}, \bibinfo {author} {\bibfnamefont {R.~J.}\ \bibnamefont {Morrison}}, \bibinfo {author} {\bibfnamefont {F.~V.}\ \bibnamefont {Murphy}},\ and\ \bibinfo {author} {\bibfnamefont {D.~E.}\ \bibnamefont {Yount}},\ }\href@noop {} {\bibfield  {journal} {\bibinfo  {journal} {Phys. Rev. D}\ }\textbf {\bibinfo {volume} {7}},\ \bibinfo {pages} {1362} (\bibinfo {year} {1973})}\BibitemShut {NoStop}%
\bibitem [{\citenamefont {Michalowski}\ \emph {et~al.}(1977)\citenamefont {Michalowski}, \citenamefont {Andrews}, \citenamefont {Eickmeyer}, \citenamefont {Gentile}, \citenamefont {Mistry}, \citenamefont {Talman},\ and\ \citenamefont {Ueno}}]{michalowski1977experimental}%
  \BibitemOpen
  \bibfield  {author} {\bibinfo {author} {\bibfnamefont {S.}~\bibnamefont {Michalowski}}, \bibinfo {author} {\bibfnamefont {D.}~\bibnamefont {Andrews}}, \bibinfo {author} {\bibfnamefont {J.}~\bibnamefont {Eickmeyer}}, \bibinfo {author} {\bibfnamefont {T.}~\bibnamefont {Gentile}}, \bibinfo {author} {\bibfnamefont {N.}~\bibnamefont {Mistry}}, \bibinfo {author} {\bibfnamefont {R.}~\bibnamefont {Talman}},\ and\ \bibinfo {author} {\bibfnamefont {K.}~\bibnamefont {Ueno}},\ }\href@noop {} {\bibfield  {journal} {\bibinfo  {journal} {Phys. Rev. Lett.}\ }\textbf {\bibinfo {volume} {39}},\ \bibinfo {pages} {737} (\bibinfo {year} {1977})}\BibitemShut {NoStop}%
\bibitem [{\citenamefont {Armstrong}\ \emph {et~al.}(1972{\natexlab{b}})\citenamefont {Armstrong}, \citenamefont {Hogg}, \citenamefont {Lewis}, \citenamefont {Robertson}, \citenamefont {Brookes}, \citenamefont {Clough}, \citenamefont {Freeland}, \citenamefont {Galbraith}, \citenamefont {King},\ and\ \citenamefont {Rawlinson}}]{armstrong1972total2}%
  \BibitemOpen
  \bibfield  {author} {\bibinfo {author} {\bibfnamefont {T.~A.}\ \bibnamefont {Armstrong}}, \bibinfo {author} {\bibfnamefont {W.~R.}\ \bibnamefont {Hogg}}, \bibinfo {author} {\bibfnamefont {G.~M.}\ \bibnamefont {Lewis}}, \bibinfo {author} {\bibfnamefont {A.~W.}\ \bibnamefont {Robertson}}, \bibinfo {author} {\bibfnamefont {G.~R.}\ \bibnamefont {Brookes}}, \bibinfo {author} {\bibfnamefont {A.~S.}\ \bibnamefont {Clough}}, \bibinfo {author} {\bibfnamefont {J.~H.}\ \bibnamefont {Freeland}}, \bibinfo {author} {\bibfnamefont {W.}~\bibnamefont {Galbraith}}, \bibinfo {author} {\bibfnamefont {A.~F.}\ \bibnamefont {King}},\ and\ \bibinfo {author} {\bibfnamefont {W.~R.}\ \bibnamefont {Rawlinson}},\ }\href@noop {} {\bibfield  {journal} {\bibinfo  {journal} {Nucl. Phys. B}\ }\textbf {\bibinfo {volume} {41}},\ \bibinfo {pages} {445} (\bibinfo {year} {1972}{\natexlab{b}})}\BibitemShut {NoStop}%
\bibitem [{\citenamefont {Skokov}\ \emph {et~al.}(2009)\citenamefont {Skokov}, \citenamefont {Illarionov},\ and\ \citenamefont {Toneev}}]{Skokov:2009qp}%
  \BibitemOpen
  \bibfield  {author} {\bibinfo {author} {\bibfnamefont {V.}~\bibnamefont {Skokov}}, \bibinfo {author} {\bibfnamefont {A.~Y.}\ \bibnamefont {Illarionov}},\ and\ \bibinfo {author} {\bibfnamefont {V.}~\bibnamefont {Toneev}},\ }\href {https://doi.org/10.1142/S0217751X09047570} {\bibfield  {journal} {\bibinfo  {journal} {Int. J. Mod. Phys. A}\ }\textbf {\bibinfo {volume} {24}},\ \bibinfo {pages} {5925} (\bibinfo {year} {2009})},\ \Eprint {https://arxiv.org/abs/0907.1396} {arXiv:0907.1396 [nucl-th]} \BibitemShut {NoStop}%
\bibitem [{\citenamefont {Bleicher}\ and\ \citenamefont {Bratkovskaya}(2022)}]{Bleicher:2022kcu}%
  \BibitemOpen
  \bibfield  {author} {\bibinfo {author} {\bibfnamefont {M.}~\bibnamefont {Bleicher}}\ and\ \bibinfo {author} {\bibfnamefont {E.}~\bibnamefont {Bratkovskaya}},\ }\href {https://doi.org/10.1016/j.ppnp.2021.103920} {\bibfield  {journal} {\bibinfo  {journal} {Prog. Part. Nucl. Phys.}\ }\textbf {\bibinfo {volume} {122}},\ \bibinfo {pages} {103920} (\bibinfo {year} {2022})}\BibitemShut {NoStop}%
\bibitem [{\citenamefont {Voronyuk}\ \emph {et~al.}(2011)\citenamefont {Voronyuk}, \citenamefont {Toneev}, \citenamefont {Cassing}, \citenamefont {Bratkovskaya}, \citenamefont {Konchakovski},\ and\ \citenamefont {Voloshin}}]{Voronyuk:2011jd}%
  \BibitemOpen
  \bibfield  {author} {\bibinfo {author} {\bibfnamefont {V.}~\bibnamefont {Voronyuk}}, \bibinfo {author} {\bibfnamefont {V.~D.}\ \bibnamefont {Toneev}}, \bibinfo {author} {\bibfnamefont {W.}~\bibnamefont {Cassing}}, \bibinfo {author} {\bibfnamefont {E.~L.}\ \bibnamefont {Bratkovskaya}}, \bibinfo {author} {\bibfnamefont {V.~P.}\ \bibnamefont {Konchakovski}},\ and\ \bibinfo {author} {\bibfnamefont {S.~A.}\ \bibnamefont {Voloshin}},\ }\href {https://doi.org/10.1103/PhysRevC.83.054911} {\bibfield  {journal} {\bibinfo  {journal} {Phys. Rev. C}\ }\textbf {\bibinfo {volume} {83}},\ \bibinfo {pages} {054911} (\bibinfo {year} {2011})},\ \Eprint {https://arxiv.org/abs/1103.4239} {arXiv:1103.4239 [nucl-th]} \BibitemShut {NoStop}%
\end{thebibliography}%

\end{document}